\documentclass[a4paper]{scrartcl}
\setcapindent{0pt}
\usepackage[utf8]{inputenc}
\usepackage{amsmath,amssymb,amstext,color,bm,verbatim,graphicx,float}
\usepackage[margin=0.85in,bottom=1in,top=1in]{geometry}
\usepackage{fancyhdr}
\usepackage{lastpage}
\usepackage{cancel}
\usepackage{booktabs}
\usepackage{enumitem}
\usepackage[normalem]{ulem}

\usepackage[numbers,sort&compress]{natbib}
\begin{document}

\begin{center}
\LARGE{\textbf{Three phases of odd robotic active matter}}    
\end{center}

%%%%%% Author list and affiliations %%%%%%%
\vspace{1cm}
\normalsize
Fan Bo$^{1,5}$, Shiqi Liu$^{1,5}$, Zenghong He$^{1}$, Wyatt Joyce$^{1}$, Gregor Leech$^{1}$, Kiet Tran$^{1}$, Keilan Ramirez$^{1}$, Nicholas Boechler$^{2,3}$, Nicholas Gravish$^{2}$, Hongbo Zhao$^{1,4}$, Tzer Han Tan$^{1,6}$
\vspace{0.04cm}

\begin{itemize}[leftmargin=0.15in]
\item[$^1$] Department of Physics, University of California San Diego, La Jolla, CA, USA \vspace{-0.3cm}
\item[$^2$] Department of Mechanical Engineering, University of California San Diego, La Jolla, CA, USA \vspace{-0.3cm}
\item[$^3$] Program in Materials Science and Engineering, University of California San Diego, La Jolla, CA, USA \vspace{-0.3cm}
\item[$^4$] Department of Chemistry, University of California San Diego, La Jolla, CA, USA\vspace{-0.3cm}
\item[$^5$] These authors contributed equally and are joint first authors. \vspace{-0.3cm}
\item[$^6$] Corresponding author: tztan@ucsd.edu \vspace{-0.3cm}
\end{itemize}
%%%%%%%%%%%%%%%%%%%%%%%%%%%%%%%%%%%%%%%%%%%

\vspace{1cm}
\normalsize
\section*{Abstract}
Nonreciprocal interactions in active matter are known to generate exotic mechanical behaviors such as odd elasticity~\cite{scheibner2020odd} and odd viscosity~\cite{banerjee2017odd,soni2019odd,markovich2024nonreciprocity}. 
However, these phenomena have largely been studied in isolation, raising a fundamental question: Is there a single system that embodies these distinct regimes of odd matter and can transition between phases, establishing a unified phase diagram for nonreciprocal active matter?
To address this, we introduce a tunable robotic active matter platform, the Magnetomechanically Augmented Spinning roBotic (MASBot) collective, in which particle-level control of chirality, activity, and pairwise interactions enables access to distinct phases of odd matter. 
By continuously increasing repulsive forces relative to attractive and transverse forces, we experimentally map a transition from an odd elastic crystal to an odd viscous liquid, and then to a chiral active gas. 
We find that this latter phase forms a non-space-filling, nonreciprocal active gas stabilized by long-range hydrodynamic attractive forces, whose statistical signatures are consistent with those of a two-dimensional self-gravitating point vortex gas~\cite{chavanis2010self,chavanis2014statistical}. 
Within these phases, adjusting spinning frequency and introducing spatially patterned activity allows us to fine-tune odd mechanical responses and tailor power spectra.
Further polar and rotational symmetry breaking at the particle scale leads to novel emergent states such as phase separation and collective translation.
Together, our system provides a fundamental experimental testbed for nonequilibrium physics and establishes a blueprint for treating robotic swarms as programmable states of matter, enabling functions that range from resilient structures to adaptive swarm reconfiguration.
% Together, our system provides a versatile macroscopic platform for chiral active matter and establishes a blueprint for robotic materials that exploit nonreciprocity for tasks ranging from resilient structures to adaptive swarm reconfiguration.

\section*{Main text}
%%%%%%%%%% Main article %%%%%%%%%%

Symmetry breaking governs the emergence of ordered phases in many-body systems, from colloidal crystals to superconductors~\cite{Anderson393,Marchetti:2013bp,bowick2022symmetry}. 
Active matter, composed of units that continuously inject energy at the microscopic scale and thereby break time-reversal symmetry, extends this paradigm by generating nonequilibrium phases absent in passive materials~\cite{bowick2022symmetry,Marchetti:2013bp}. 
Studies of polar and nematic active fluids in biological and synthetic systems have revealed phenomena such as flocking~\cite{toner1998flocks,julicher2018hydrodynamic}, active turbulence~\cite{alert2022active,dunkel2013fluid,Tan:2020jk}, and topology-driven dynamics~\cite{shankar2022topological}. 
Beyond polar and nematic order, chiral activity provides a distinct route to nonequilibrium organization, in which intrinsic rotation and handed interactions yield parity-violating stresses~\cite{fruchart2023odd,scheibner2020odd,soni2019odd,banerjee2017odd} and edge currents~\cite{soni2019odd}. 
A unifying feature across these systems is the role of nonreciprocal interactions, which produce odd mechanical response, sustained wave propagation, and non-Hermitian dynamics~\cite{soni2019odd,banerjee2021active,banerjee2017odd,scheibner2020odd,fruchart2023odd,souslov2019topological}. \\

Nonreciprocal behavior in active matter spans natural chiral systems \cite{tan2022odd,chao2024selective} and a range of synthetic platforms \cite{soni2019odd,bililign2021chiral,veenstra2025adaptive}.
While these systems have revealed that nonreciprocal interactions can generate odd elasticity \cite{scheibner2020odd,baconnier2022selective} or odd viscosity~\cite{banerjee2017odd,soni2019odd,markovich2024nonreciprocity}, they have been restricted to a single manifestation of such behavior, \textit{e.g.} odd elasticity \textit{or} odd viscosity, but not both. 
Collectives of robots are an emerging platform that have revealed new physics of active matter systems \cite{goldman2024robot, wang2017dynamic,Jaeger2024ScienceRobotics,veenstra2025adaptive,levine2023physics,savoie2019robot}. In this work, we show that robots can introduce and tune nonreciprocal interactions that are difficult to realize in ``natural'' systems, enabling odd matter with multimodal, qualitatively tunable behaviors and opening experimental access to new physical regimes.

\begin{figure}[]
\centering
\small
\includegraphics[width=0.9\textwidth]{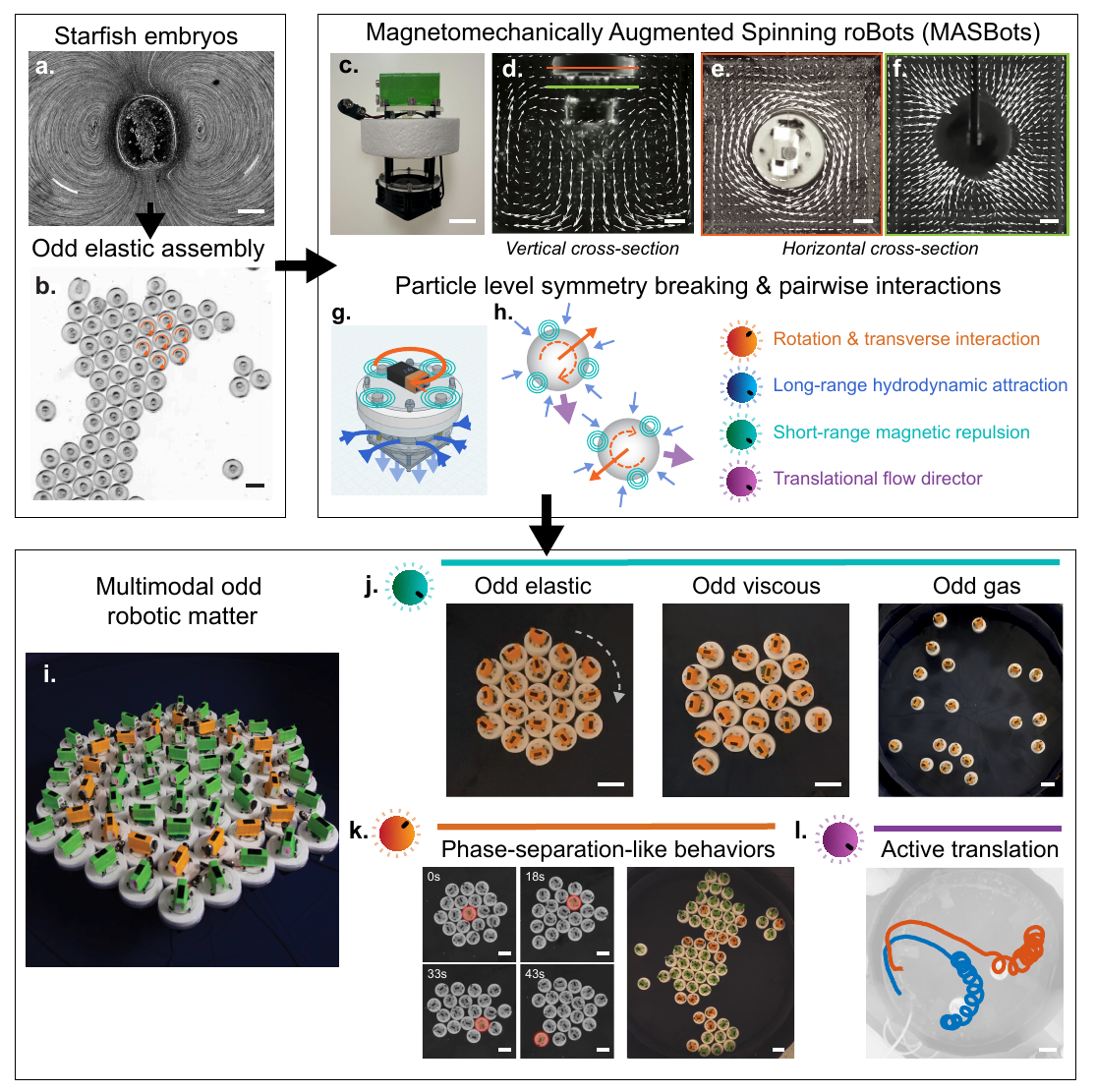}
\caption{\textbf{Magnetomechanically Augmented Spinning roBotic (MASBot) collectives.} 
\textbf{a-c.}~Inspired by starfish embryos that self-assemble into living chiral crystals (a-b), we build MASBot composed of a cylindrical base with a rotating propeller (c). Scale bars are $100~\mathrm{\mu m}$ and $200~\mathrm{\mu m}$ in (a) and (b), respectively. 
\textbf{d-f.} Single MASBot shows vortical flow in the vertical plane (d), clockwise rotational flow in the top horizontal plane (e), and inward attractive flow at the mid-horizontal plane (f). Scale bars in (c-f) are $2~\mathrm{cm}$. 
\textbf{g.}~Rotational flow exerts nonreciprocal transverse and attractive interactions, and magnets induce short-range magnetic repulsion on neighboring MASBots. \textbf{h.}~Activity and pairwise interactions can be tuned at single MASBot level, including translational self-propulsion.
\textbf{i-l}~MASBot collectives exhibit dynamic reorganization, forming clusters with hexagonal symmetry due to hydrodynamic attraction (i). With increasing short-range magnetic repulsion, the collective undergoes phase transitions from solid to gas (j). By tuning the rotation speed, MASBot collectives show phase-separation-like behaviors (k). Introducing a translational flow director to a standard MASBot breaks polar symmetry and enables the MASBots to actively translate (l). Scale bars in (j-l) are one MASBot  diameter ($7.5~\mathrm{cm}$).}
\label{fig:intro}
\end{figure}

\subsection*{Self-assembly of Magnetomechanically Augmented Spinning roBotic (MASBot) collective}

Inspired by the spontaneous self-assembly of starfish embryos into living chiral crystals~\cite{tan2022odd} (Fig.~\ref{fig:intro}a-b), we introduce a type of internally driven matter---which we name as Magnetomechanically Augmented Spinning roBot (MASBot) collectives---that can tunably break multiple symmetries at particle level, allowing multimodal transitions between qualitatively distinct phases of odd behavior. 
In their simplest form, each MASBot consists of a cylindrical base that floats on the water surface with a rotating propeller underneath that generates rotational torque (breaking chiral symmetry), nonreciprocal transverse forces and a long-range radial hydrodynamic attractive force (Fig.~\ref{fig:intro}c and Methods~1.1-1.2). Fixing magnets onto MASbot surface provides a repulsive force that dominates at short-range.
Similar robot collectives have allowed for studying controllable attractive interactions \cite{li2021programming}, nonreciprocal chiral interactions in low Reynolds number regimes \cite{gelvan2025hydrodynamic}, and rotational contact-based interactions \cite{scholz2018rotating}. 
However, no prior platforms have enabled control of all three modes of interaction with robots that move in inertial regimes.  
Hydrodynamic imaging around a single MASBot (SI Video~1 and Methods~1.3) shows a pair of counter-rotating vortices in the vertical plane (Fig.~\ref{fig:intro}d), indicating a toroidal vortex structure in 3D due to axial symmetry, which is reminiscent of the flows around starfish embryos (Fig.~\ref{fig:intro}b,~\cite{tan2022odd}), but five orders of magnitude larger in scale at Reynolds number of Re $\sim 4000$. 
Horizontal cross-sections show clockwise rotation, inducing nonreciprocal transverse active interactions on neighbors (Fig.~\ref{fig:intro}e), and net inward flow, driving attractive interactions between MASBots (Fig.~\ref{fig:intro}f).\\

A central advantage of this design is the tailorability: multiple symmetries can be controllably broken at the particle level (\textit{i.e.}, single MASBot), including chiral (transverse) symmetry via spinning speed and direction, radial symmetry through placement of repulsive magnetic dipoles around the MASBot, and polar symmetry via addition of a flow director (Fig.~\ref{fig:intro}h). 
Of particular note, the hydrodynamic-mediated transverse interaction is nonreciprocal~\cite{tan2022odd}, in contrast to some chiral active matter systems where chiral bias appears in the self-propulsion term (\textit{e.g.}, Ref.~\cite{ma2022dynamical}), such that the MASbots behave similarly to point vortices in the 2D Onsager model that get advected by the circulation of their neighbors~\cite{OnsagerRevModPhys}.\\

Herein, we observe that MASBots self-organize into active solid and liquid states with parity-violating odd mechanical responses, and active gas states accessible through inertial effects (Fig.~\ref{fig:intro}j). We find that the chiral active gas phase is particularly remarkable, as, via the combination of inertial dynamics, non-space-filling structure set by long range hydrodynamic attraction, and nonreciprocal, parity violating correlations, it forms a direct analogy with a self gravitating two-dimensional (2D) point vortex gas~\cite{chavanis2010self,chavanis2014statistical}.
Further, we find that tunability at the particle level enables heterogeneous active ensembles to exhibit spatially patterned active driving and phase separation (Fig.~\ref{fig:intro}k and SI Videos~2, 12). Finally, breaking polar symmetry via inclusion of a flow director on the MASBot allows for hydrodynamic alignment and directed motion (Fig.~\ref{fig:intro}l).
In contrast to robotic collectives that rely on sensing or centralized control, MASBot collectives exploit intrinsic chiral interactions to generate emergent, self-organizing dynamics with tunable properties. 
By bridging theory and experiment across scales, this platform establishes a new class of active matter that reshapes how emergent phases of driven systems can be studied and engineered.

\begin{figure}[h!]
\centering
\small
\includegraphics[width=1\textwidth]{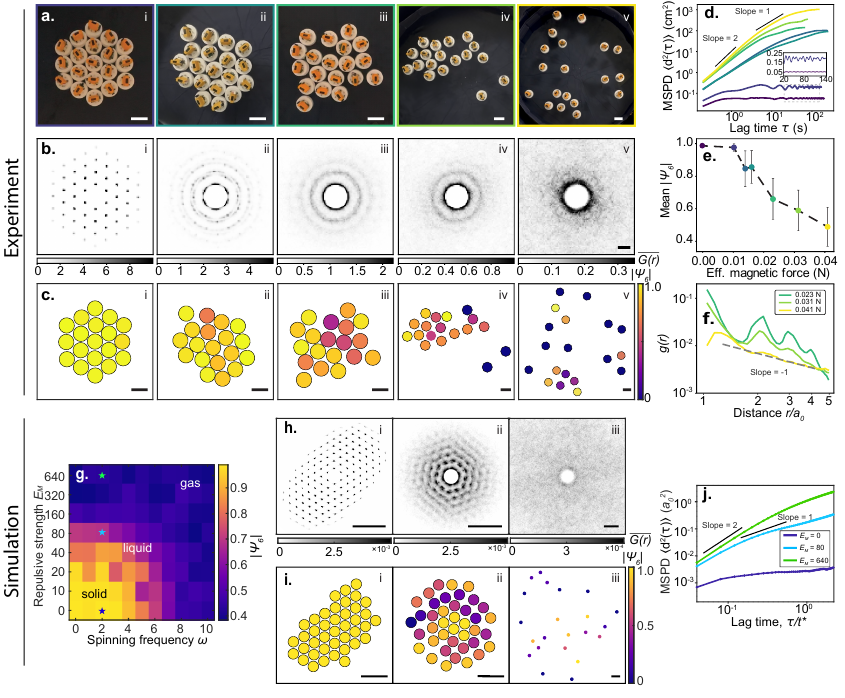} 
\caption{\textbf{Phase transitions in chiral active matter.} 
\textbf{a-f.}~Experimental analysis of phase transitions due to increasing magnetic repulsion. (a) Snapshots of MASBot collectives, from the solid to gas state, with correspondingly increasingly stronger magnetic repulsive force (left to right, i-v). Time averaged 2D pair distribution function $G(\textbf{r})$ (b) and the local bond orientational order parameter $\psi_6$ (c) for the corresponding area in (a). Scale bars in (a-c) are $7.5~\mathrm{cm}$.
Mean squared pairwise displacement (MSPD) as a function of lag time $\tau$ for different states from solid (purple) to gas (yellow) (d). The inset is the oscillation seen in the solid states (linear scale). (e) Mean $\psi_6$ as a function of magnetic repulsion. 
(f) Radial distribution function $g(r)$ for liquid ($0.023~\mathrm{N}$), liquid-gas ($0.031~\mathrm{N}$), and gas ($0.041~\mathrm{N}$) states.
\textbf{g-j.}~Numerical simulations of phase transitions.
Phase diagrams of $|\psi_6|$ (g) as functions of spinning frequency $\omega$ and repulsive strength $E_M$. (h) Time averaged $G(\textbf{r})$, shown from left to right for the solid ($E_M=0$), liquid ($E_M=80$), and gas states ($E_M=640$) marked in (g) where $\omega = 2$. (i) $\psi_6$ corresponding to (h). (j) MSPD as a function of $\tau$ for the states in (h, i). Scale bars in (h, i) are 2.5 MASBot diameters.}
\label{fig:phase}
\end{figure}

\subsection*{Observation of phase transitions in chiral robotic matter}
We first investigate nonequilibrium phase transitions accessible in MASBot collectives by tuning the strength of the magnetic repulsion (by changing the number of magnets around the MASBot's circumference). 
Without magnetic repulsion, radial interactions remain purely attractive (through hydrodynamics), and the collective self-assembles into robotic chiral crystals that undergo rigid-body rotation. 
The cluster forms a highly ordered hexagonal lattice (Fig.~\ref{fig:phase}a(i)) with strong translational order as quantified by the 2D pair distribution function $G(r)$ (shown by the prominent hexagonal peaks in Fig.~\ref{fig:phase}b(i)), and orientational order as quantified by the bond orientational order parameter $\psi_6$ (shown by $\psi_6\sim1$ in Fig.~\ref{fig:phase}c(i)).
Dynamically, we observe a flat mean-squared pairwise displacement (MSPD), a signature of the solid phase (Fig.~\ref{fig:phase}d). The MSPD displays prominent spontaneous oscillations at long time scale (Fig.~\ref{fig:phase}d, inset), which is indicative of strain waves predicted for odd elastic solids. \\

As magnetic repulsion increases, the system fluidizes (Fig.~\ref{fig:phase}a(ii-iv) and SI Video~3-5). 
Structurally, $G(r)$ retains signatures of sixfold symmetry, but density becomes diffuse due to nearest-neighbor exchange (Fig.~\ref{fig:phase}b(ii-iv), SI Fig.~4c, and Methods 2.2). 
We observe larger $\psi_6$ fluctuations (Fig.~\ref{fig:phase}c(ii-iv) and Methods 2.3), while the MSPD acquires a slope of $\sim 1$ (Fig.~\ref{fig:phase}d and Methods 2.4), consistent with liquid-like transport. Moreover, analysis of translational and orientational correlation functions reveals signatures of a hexatic phase (Fig.~\ref{fig:phase}a(ii-iii), SI Fig.~5, and Methods 2.5-2.6), consistent with both experimental results on ferrofluid droplets~\cite{guillet2025melting} and predictions from Kosterlitz–Thouless–Halperin–Nelson–Young (KTHNY) theory~\cite{halperin1978theory,kosterlitz1973ordering}, which proposes that the transitions between the crystalline and liquid phases are mediated by thermally activated topological defects. This fluidized regime shows signatures of odd viscosity (discussed later), establishing that MASBotS can access multiple odd mechanical responses within a single system.\\

At high repulsion, the ensemble transitions into a dilute gas (Fig.~\ref{fig:phase}a(v)). Due to the long-range hydrodynamic attraction balanced by strong repulsion, we observe non-space-filling configurations, diffuse $G(r)$ (Fig.~\ref{fig:phase}b(v)) with only a single ring, and power-law radial distribution tails $g(r)\sim r^{-1}$ (Fig.~\ref{fig:phase}c(v),f). Because local hexagonal order is lost in this regime, the bond orientational order parameter $\psi_6$ is no longer strictly meaningful; 
we nevertheless compute $\psi_6$ to quantify the disordered nature of the gas (Fig.~\ref{fig:phase}c(v)). The MSPD becomes superdiffusive with a slope close to 2 (Fig.~\ref{fig:phase}d), indicating the onset of inertial transport. This inertial, non-space-filling gas displays pair statistics analogous to a 2D self-gravitating gas~\cite{chavanis2010self,chavanis2014statistical} (power-law decay of $g(r)$).\\

To elucidate the underlying mechanics, we build a coarse-grained inertial spinner simulation model that incorporates long-range hydrodynamic attraction, short-range magnetic and Weeks–Chandler–\\Andersen (WCA) repulsion, and transverse nonreciprocal interaction that encodes chiral interactions (Methods~3). 
Considering $\psi_6$ (Fig.~\ref{fig:phase}g), the simulated phase diagram confirms that increasing either the repulsive force ($E_M$) or the spinning frequency $\omega$ (which governs the transverse force) leads to fluidization and a corresponding decrease in orientational order. 
Tuning the repulsion alone achieves solid, liquid, and gas states in simulation (Fig.~\ref{fig:phase}h-i and SI Fig. 6, 7) with $G(r)$ and bond orientational order parameter $\psi_6$ consistent with experiments. Notably, the ballistic MSPD scaling in the gas phase emerges only when inertia is retained (Fig.~\ref{fig:phase}j and SI Fig.~8), underscoring the role of inertia in the emergence of this self-attracting nonreciprocal gas phase.\\

Together, the experiment–simulation comparison shows that MASBots provide the first macroscopic system that can be tuned continuously across an odd elastic solid, an odd viscous liquid, and a nonreciprocal active gas. This continuous, multimodal control sets the stage for the detailed characterization of odd viscosity (Fig. 3) and odd elasticity (Fig. 4) in the following sections.

\begin{figure}
\centering
\small
\includegraphics[width=1\textwidth]{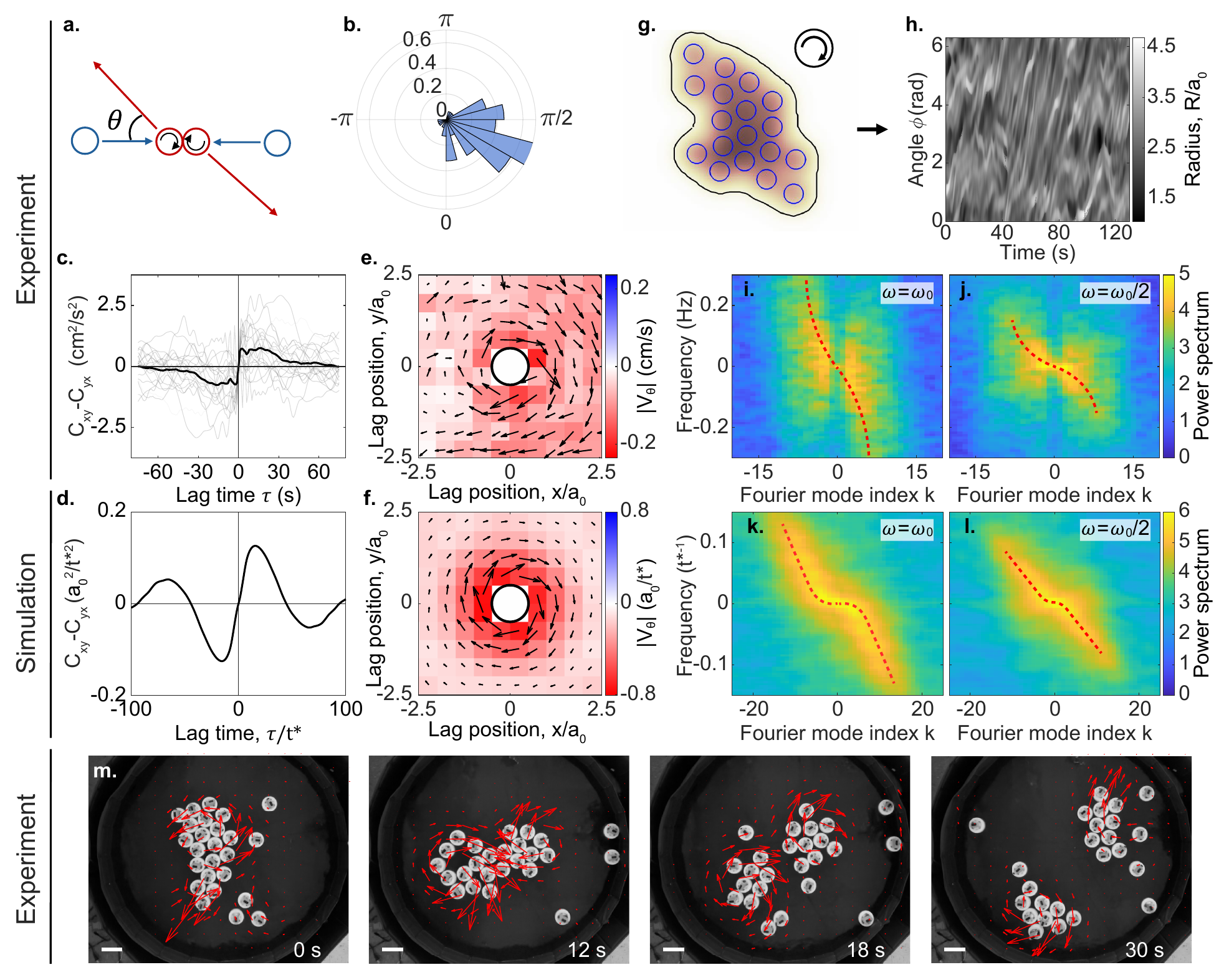}
\caption{\textbf{Odd viscosity in chiral robotic matter extending to active chiral gas enabled by inertia.} 
\textbf{a-b.}~Schematic (a) and scattering angle distribution (b) of collision experiment, indicating microscopic parity violation.
\textbf{c-d.}~The temporal velocity correlation of gas state $C(\tau)$ in experiment (c) and simulation (d). \ The gray lines in (c) are single particle $C(\tau)$ and the black line is the averaged $C(\tau))$.
\textbf{e-f.} The separation conditioned velocity correlation field $\mathbf{V}(\mathbf{r})$ in experiment (e) and simulation (f). The arrows indicate the direction and magnitude of the velocity. The color map in the background shows the magnitude of the angular component of velocity $V_{\theta}$. The black circle in the center indicates the particle size.
\textbf{g-h.}~Schematic of the boundary $R(\phi,t)$ (Black boundary) of MASBot collectives (g), and the angular-time kymograph (h).
\textbf{i-j.}~Power spectrum of angular-time kymograph shows spinning frequency dependent parity breaking in experiment for full frequency $\omega_0$ (g) and half frequency $\omega_0/2$ (h). The red dashed line (guide for the eye) highlights the peak in the spectrum.
\textbf{k-l.}~Power spectrum of angular-time kymograph shows spinning frequency dependent parity breaking in simulation for full frequency $\omega_0$ (i) and half frequency $\omega_0/2$ (j). The red dashed line (guide for the eye) highlights the peak in the spectrum. The parameters in simulation are $\omega = 20, E_M = 160$ for (d) and (f) (gas phase), and $\omega_0 = 6, E_M = 80$ for (h) (liquid phase). \textbf{m.}~Spontaneous droplet fission driven by odd instabilities in a MASBot cluster with effective magnetic repulsion $ 0.016~\mathrm{N}$ (liquid phase, also see SI Video 9). Scale bar: $10~\mathrm{cm}$.} 
\label{fig:oddviscosity}
\end{figure}

\subsection*{Odd stresses and parity violation in active chiral robotic fluid}
Next, we investigate how microscopic chirality generates parity-violating stresses in the MASBot fluid (liquid and gas) states. 
We first isolate the two-body contribution by performing pairwise scattering experiments (Fig.~\ref{fig:oddviscosity}a and Methods~1.5). 
When two spinning MASBots collide, each exerts a handed transverse force on the other, producing a strongly asymmetric distribution of scattering angles (Fig.~\ref{fig:oddviscosity}b). This macroscopic demonstration of chiral scattering, typically confined to colloidal or molecular scales~\cite{reichhardt2019active,han2021fluctuating,reichhardt2022active}, establishes the elementary nonreciprocal interaction underlying the fluid’s odd response.\\

To probe collective parity violation, we analyze the chiral temporal velocity autocorrelation~\cite{han2021fluctuating} $C(\tau)=\langle\mathbf{v}(t+\tau)\times \mathbf{v}(t)\rangle_t = \langle v_x(t+\tau)\,v_y(t)\rangle_t - \langle v_y(t+\tau)\,v_x(t)\rangle_t = C_{xy}(\tau)-C_{yx}(\tau)$ (Methods~2.8). We find that in experimental gas state (Fig.~\ref{fig:oddviscosity}c and SI Fig. 9a-f), the correlation $C$ breaks time reversal symmetry $C(\tau)\neq C(-\tau)$ and shows parity violation $\mathcal{P}[C(\tau)] = -C(\tau) \neq C(\tau) $. However, the overall correlation $C$ remains PT symmetric $C(\tau) = -C(-\tau)= \mathcal{P}[C(-\tau)])$ at steady state, which is consistent with the fact that reversing time is equivalent to reversing the spinner chirality in our system. However, we note that when transient dynamics are considered, this PT symmetry no longer holds. For example, if we play a video of the coalescing collective backwards, it does not match the dynamics of a coalescing collective with opposite MASBot rotation direction. We further define the separation-conditioned velocity distribution function $\mathbf{V}(r)=\left\langle\, \mathbf{v}_j \,\middle|\, r_{ij}=r \right\rangle_{i,j}$ (intuitively, this reports the average velocity of neighboring MASBots at separation $r$, see Methods~2.7), which shows circulating particle flow and pronounced clockwise angular components $V_{\theta}$ (Fig.~\ref{fig:oddviscosity}e and SI Fig. 10a-f), revealing the chiral nature of transverse interaction at microscopic level, consistent with dynamics of Onsager model of same-signed point vortices. Identical analysis in simulations of the gas phase reproduce these signatures (Fig.~\ref{fig:oddviscosity}d, f and SI Fig. 9g-l, 10g-l, 8). Altogether, this shows that the gas phase, in addition to being non space filling and inertial (as per Fig.~\ref{fig:phase} and SI Fig. 8), is also parity violating, such that it can be considered akin to a self-attracting 2D point vortex gas~\cite{chavanis2010self,chavanis2014statistical}, which is a previously unobserved regime of chiral active matter.\\

Next, we investigate the odd dynamics in an active chiral robotic liquid.
Generically, odd stresses arising from rotational viscosity produce antisymmetric components of the stress tensor and can drive unidirectional boundary waves~\cite{fruchart2023odd,banerjee2017odd,soni2019odd}.
To characterize this effect, we measure the boundary fluctuations of MASBot collectives $R(\phi,t)$ and perform a Fourier transform to extract the power spectrum~\cite{soni2019odd} (Fig.~\ref{fig:oddviscosity}g, h and Methods~2.9). 
The resulting power spectrum is strikingly asymmetric (Fig.~\ref{fig:oddviscosity}i), revealing parity-violating spectral weight. 
Halving the spinning frequency decreases the slope of the power spectrum but retains the asymmetry (Fig.~\ref{fig:oddviscosity}j), demonstrating that odd hydrodynamic response is tunable at the particle level.
Such spectral asymmetry in a small-N inertial system represents a new experimental window into odd hydrodynamics, extending the concept beyond colloidal realization~\cite{soni2019odd}. 
Our simulation results in a liquid regime which also captures the spinner's frequency-dependent, asymmetric power spectra (Fig.~\ref{fig:oddviscosity}k, l).\\

Another distinctive feature of odd chiral liquids is the emergence of odd instabilities. 
Rotational viscosity generates antisymmetric odd stresses, which in turn drive unidirectional edge currents. When edge currents on opposite interfaces overlap due to fluctuations, their interaction can give rise to unstable modes, ultimately leading to the breakup of fluid clusters~\cite{soni2019odd}. In our experiments, we observe similar signatures during droplet breakup events (Fig.~\ref{fig:oddviscosity}m, SI Video~9), marking the first observation of such odd hydrodynamic instabilities in an engineered robotic liquid. This further confirms the presence of antisymmetric odd stresses and chiral edge dynamics in the MASBot liquid.\\

\begin{figure}[h!]
\centering
\small
\includegraphics[width=0.76\textwidth]{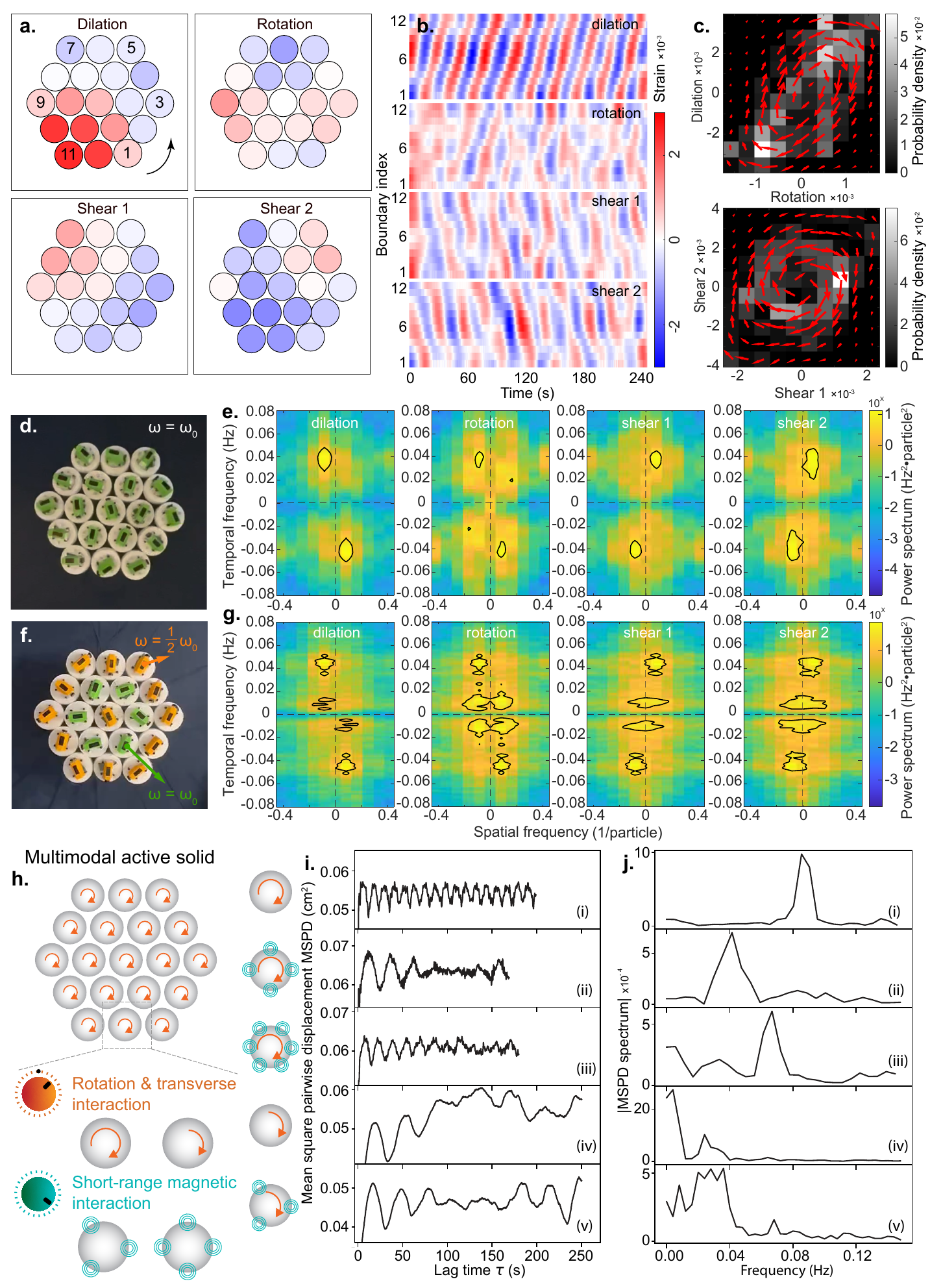}
\caption{\textbf{Odd waves and cycling states in nonreciprocal active solid.} 
\textbf{a-c.}~Spontaneous strain waves and odd elastic engine cycle (Methods~2.10-2.11). A snapshot of the MASBot collective decomposed into the four principal strain components (a). Space-time kymographs of strain waves along the boundary (counterclockwise) of a MASBot solid (b). Probability density current in the dilation–rotation strain (top) and shear 1–shear 2 strain (bottom) shows counterclockwise strain cycle (c).
\textbf{d-g.}~Strain power spectra of homogeneous (d-e) and  patterned (f-g) active solid. Active solid with two layered spinning frequencies (f, $\omega_0=2.63~\mathrm{Hz}$) exhibits secondary peaks (g) in the power spectra where the peaks are marked with black lines and defined as $60\%$ of the highest value.
\textbf{h-j.}~Excitable frequencies modulated by rotation speed (orange arrows) and p-atic symmetry of repulsion (green circles). Tuning the particle-level interactions (h) excites distinct vibrational frequencies (i-j). Case (ii) corresponds to the case in Fig.~\ref{fig:activesolids}d, while cases (i) and (iii) correspond to the cases in Fig.~\ref{fig:phase}d with magnetic repulsion of $0~\mathrm{N}$ and $0.01~\mathrm{N}$ respectively.}
\label{fig:activesolids}
\end{figure}

\subsection*{Spontaneous strain waves and mode coupling reveal signatures of odd elasticity}

Next, we investigate odd dynamics in the solid state. Odd elasticity theory~\cite{scheibner2020odd,choi2024noise}, predicts that nonreciprocal solids exhibit spontaneous strain waves and energy-cycling deformation modes, and has been reported in living chiral crystals of starfish embryos~\cite{tan2022odd,chao2024selective} and in active synthetic materials~\cite{baconnier2022selective,veenstra2025adaptive}. Consistent with this, our earlier mean squared pairwise displacement (MSPD) measurements (Fig.~\ref{fig:phase}d) revealed signatures of spontaneous oscillations in MASBot particle displacements within the solid state. \\

To quantify these wave dynamics more directly, we track the positions of MASBot particles in the co-rotating frame, compute the displacement field, and extract the spatiotemporal patterns of the four principal strain components: dilation, rotation, ``shear 1'', and ``shear 2'' (Fig.~\ref{fig:activesolids}a and Methods~2.10), as defined in~\cite{scheibner2020odd}. We find that all four strain fields exhibit robust wave propagation throughout the active solid (Fig.~\ref{fig:activesolids}b and SI Video 10). To visualize these dynamics, we focus on strain wave patterns along the boundary of the solid cluster (as indicated by boundary indices in Fig.~\ref{fig:activesolids}a) and plot the space-time kymographs for the four components (Fig.~\ref{fig:activesolids}b). We find that the dilation strain propagates as highly regular counterclockwise unidirectional waves, while shear 1 and shear 2 propagate clockwise. This chiral symmetry breaking is further corroborated by the power spectra (Fig.~\ref{fig:activesolids}d-e), which display prominent parity-violating peaks: top-left/bottom-right quadrants for dilation strain, and top-right/bottom-left quadrants for shear 1 and shear 2. Further, by constructing the probability current associated with the joint strain distribution and examining its circulation in strain–strain phase space~\cite{battle2016broken,tan2022odd,chao2024selective} (see Methods~2.11), we find closed, handed cycles linking dilation–rotation and shear 1–shear 2 (Fig.~\ref{fig:activesolids}c), demonstrating continuous energy flow through orthogonal deformation modes, as predicted by odd elasticity theory~\cite{scheibner2020odd} and consistent with the odd cycles observed in starfish embryo living chiral crystal~\cite{tan2022odd,chao2024selective}.\\

A distinctive feature of the MASBot platform is the ability to tune microscopic interactions and create multimodal active solids. By programming alternating lattice layers to spin at different frequencies (Fig.~\ref{fig:activesolids}f), we impose a patterned nonequilibrium drive that induces secondary spectral peaks in the strain-wave power spectrum (Fig.~\ref{fig:activesolids}g). This demonstrates that odd-elastic excitations can be structured through particle-level programming.\\

Finally, we revisit the spontaneous oscillations observed in Fig.~\ref{fig:phase}d. 
By independently tuning the transverse nonreciprocal interaction strength through the spinning frequency $\omega$ and adjusting the $p$-atic symmetry ($p$-fold rotational symmetry) of repulsion via magnet placement (orange arrows and green circles in Fig.~\ref{fig:activesolids}h), we directly control the symmetry and amplitude of odd-elastic forcing. 
We note that these control parameters are not strictly independent: varying spinning frequency $\omega$ simultaneously affects both the strength of nonreciprocal interactions and the characteristic frequency of odd forcing, while changes in magnet placement change both the strength and rotational symmetry of repulsion strength.
Quantifying these oscillations using the MSPD reveals distinct, tunable frequency content for different $\omega$ and $p$-atic configurations (Fig.~\ref{fig:activesolids}i-j), demonstrating that the excitable modes of the odd elastic solid are programmable at the particle level. 
This tunability shows that the MASBot solid not only supports spontaneous odd-elastic waves but also allows their spectral properties to be reshaped through controlled manipulation of microscopic chirality and interaction symmetry.\\

\begin{figure}[h!]
\centering
\small
\includegraphics[width=0.8\textwidth]{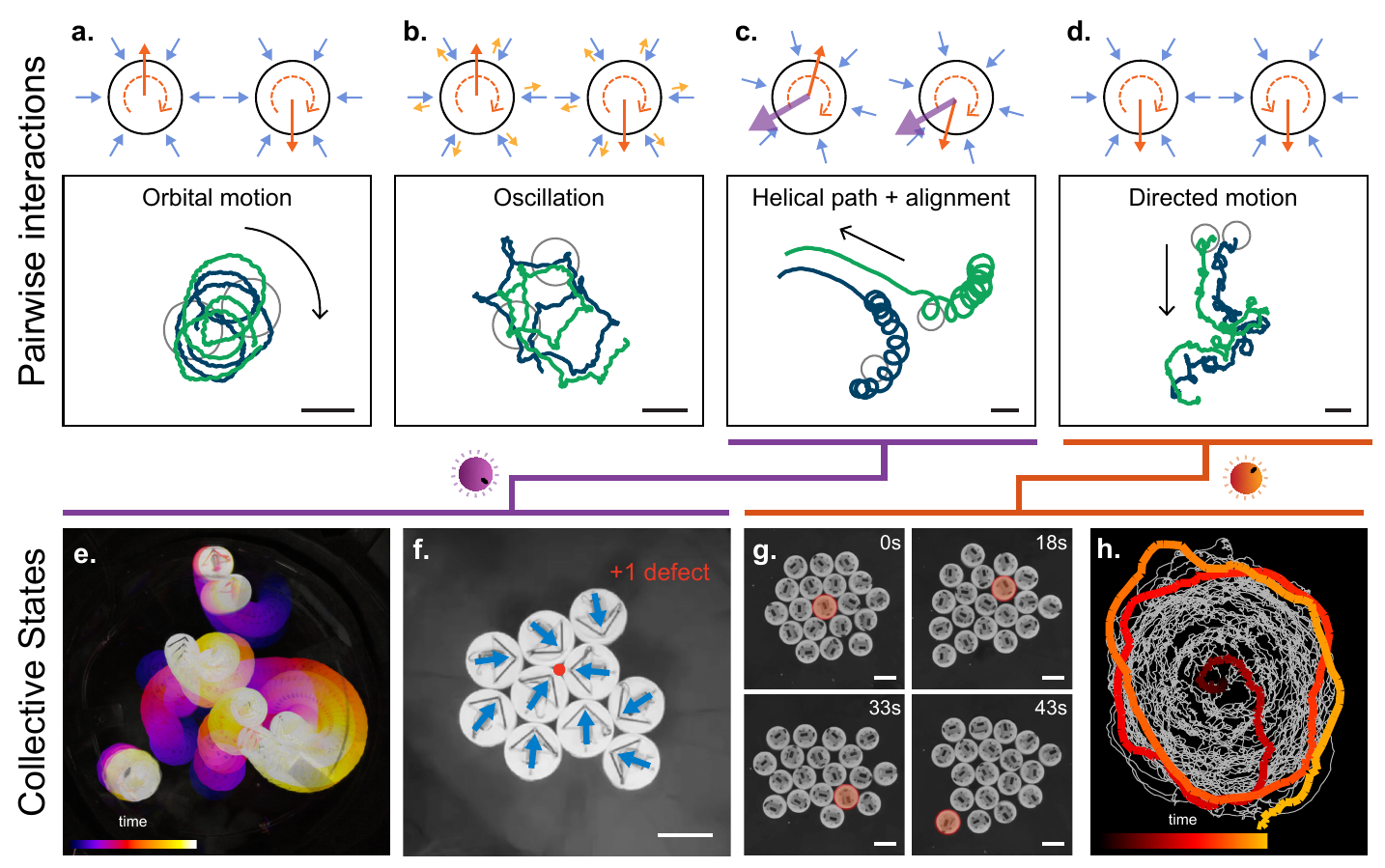}
\caption{\textbf{Tunable interactions through broken symmetries.} 
\textbf{a.}~A pair of MASBot particles shows orbital motion under hydrodynamic attraction and transverse interaction.
\textbf{b.}~Shallow water introduces short range repulsion, $r$, resulting in oscillatory nonlinear dynamics. 
\textbf{c.}~By attaching a slat to redirect fluid flow, MASBot particle breaks polar symmetry, resulting in helical path and neighbor alignment.
\textbf{d.}~Two MASBot particles with opposite chirality results in net directed motion. Scale bars in (a-d) and (g): $7.5~\mathrm{cm}.$
\textbf{e-f.}~Collective of self-propelled MASBot particles shows dynamic reorganization and eventually settle into a rigid body when a $+1$ topological defect (red dot in (f), blue arrow: the propelling direction) emerge at center of cluster. The scale bar in (f): $12.5~\mathrm{cm}$.
\textbf{g-h.}~A faster spinning MASBot can robustly escape an ensemble of slower spinning particles. 
}
\label{fig:addons}
\end{figure}

\subsection*{Tunable interactions through broken symmetries}
Thus far, we have shown how MASBots enable continuous tuning of collective behavior by adjusting repulsion and rotation rate, thereby accessing solid, liquid, and gas phases and generating both odd viscosity and odd elasticity. We now extend this multimodality to demonstrate that additional microscopic symmetries: chirality, polar symmetry, and interaction anisotropy, can also be broken deliberately to engineer new interaction rules. Because MASBots combine hydrodynamic attraction with transverse nonreciprocal interactions, even a pair of MASBots exhibits rich dynamics, including orbital motion (Fig.~\ref{fig:addons}a) and oscillatory radial modes which arise from the interplay between 3D hydrodynamic flow, inertia and short-range repulsion (Fig.~\ref{fig:addons}b). These behaviors emerge without sensing or communication, emphasizing that microscopic broken symmetries are sufficient to generate new effective interactions.\\

Further modifications expand this design space. Adding an angled slat breaks polar symmetry and produces self-propulsion, yielding helical trajectories and alignment interactions (Fig.~\ref{fig:addons}c) that, when many MASBots interact within a dense cluster, drive the system toward ordered states before settling into a globally rotating structure characterized by a single +1 topological defect, where the local alignment direction (blue arrows, Fig.~\ref{fig:addons}f) winds once around a central point (red dot in Fig.~\ref{fig:addons}f, SI Video~11 and Methods~2.12). Altering chirality introduces additional interaction modes. For instance, opposite-handed pairs generate directed motion (Fig.~\ref{fig:addons}d)~\cite{gardi2023demand,reichhardt2020dynamics}. In addition, a high frequency spinner acting as a defect in an otherwise low frequency collective will escape and orbit around slower neighbors (Fig.~\ref{fig:addons}g-h and SI Video~12), providing a mechanism to navigate or reshape collective topology. Together, these results show that MASBots are not limited to tuning interaction strength, but can fundamentally reconfigure the type of interaction through controlled symmetry breaking.\\ 
 
\subsection*{Discussion}
Altogether, this work introduces a macroscopic active matter platform that can be tuned continuously between odd elastic solids, odd viscous liquids, and a nonreciprocal chiral active gas with vortex-gas statistics. By enabling particle-level control of chirality, interaction symmetry, and driving, MASBots provide a unified setting in which multiple odd mechanical responses emerge within a single physical system. In contrast to globally actuated systems, such as magnetically-driven colloidal spinners \cite{zhang2022polar,zhang2020reconfigurable,ceron2023programmable,soni2019odd,bililign2021chiral} and vibrated chiral grains \cite{scholz2018rotating,caprini2024dynamical}, individual addressability enables controlled heterogeneous and spatially patterned collectives, in which different particles can be programmed with distinct magnitudes and types of ``oddness'', allowing systematic exploration of mixed odd phases and composite symmetry-broken states. Further, because of the particle level tunability, we suggest that, in the future, sparse control strategies, paired with the physically emergent dynamics, could be used to actively control the behavior of the MASBot collective (e.g.~\cite{nguyen2014emergent,spellings2015shape}). Our results also demonstrate that controlled symmetry breaking can be used to sculpt phases, excitations, and instabilities in driven many-body systems, allowing experimental tests of concepts that have remained largely theoretical, such as informational active matter~\cite{vansaders2023informational}. At the same time, the MASBot platform offers a new direction for robotics, where collective functions arise from designed interactions rather than centralized control or communication. This approach points toward robotic materials that exploit physical nonreciprocity, tunable coupling, and emergent organization to achieve functions that are difficult to realize with conventional algorithms.

\subsection*{Acknowledgments}
We thank Zhi-Feng Huang, Daniel Parker, Vishal Patil and Yu-Chen Chao for their feedback to manuscript. This work was supported by the Naval Innovation, Science, and Engineering Center under ONR grant N000142312831. N.B. acknowledges support from the US Army Research Office (Grant No. W911NF-20-2-0182). T.T. acknowledges support from UC MRPI Active Matter Hub.\\

\subsection*{Author contributions}
N.B., N.G., H.B. and T.T.H. designed the research. 
F.B., Z.H., W.J., G.L., K.T. and K.R. performed the measurements. 
F.B., S.L., Z.H., W.J., G.L., K.R. and T.T.H. analyzed the results. 
S.L. and H.Z. performed and analyzed the numerical simulations.  
All authors discussed the results and co-wrote the paper.

\subsection*{Competing interests}
The authors declare no competing interests.

% \printbibliography
\bibliography{references}

\end{document}